\documentstyle[aps]{revtex}

\begin{document}
\author{Qi Li$^1$, Yong Chen$^{1,2}$ and Ying Hai Wang$^1$}
\title{Coupling parameter in synchronization of diluted neural networks\thanks{%
Published in Phys. Rev. E65, 041916(2002)}}
\address{$^1$Department of Physics, Lanzhou University, Gansu, 730000, China\\
$^2$State Key Laboratory of Frozen Soil Engineering, CAREERI, CAS, Lanzhou,\\
730000, China}
\maketitle

\begin{abstract}
We study the critical features of coupling parameter in the synchronization
of neural networks with diluted synapses. Based on simulations, the
exponential decay form is observed in the extreme case of global coupling
among subsystems and fully linking in each network: there exists maximum and
minimum of the critical coupling intensity for synchronization in this
spatially extended system. For the partial coupling, we present the primary
result about the critical coupling fraction for various linking degrees of
networks.
\end{abstract}
\bigskip

Synchronization of coupled complex systems has been an intensively studied
subject since the pioneering work of Fujisaka\cite{fujisaka 1983} and others%
\cite{pecora 1990}. This phenomenon of synchronization is observed in many
other fields, such as in neural networks\cite{abarbanel 1996}, in biological
populations\cite{winfree 1980} and in chemical reactions\cite{khrustova 1995}%
. Recently, spatially extended system has inspired great interest\cite
{morelle 1998}.

Following the series of work contributed by Zanette\cite{morelle 1998}\cite
{morelli 2001}\cite{zanette 1998}, we consider a simple modified version of
the neural network model described in \cite{zanette 1998}. As a very
important ubiquitous factor corresponding to real biotic neural systems, the
neural activity and morphology of synaptic connectivity i.e. the dilution of
neural networks must be introduced\cite{kitano 1998}. That is, in natural
neural systems, not all the neurons are linked together. So, there is a
chance to investigate the critical features of coupling parameter and the
function of structural topology in synchronization of extended systems.

We consider a neural network model that consists of $N$ analog neurons $%
x_i\left( t\right) \in \left[ 0,1\right] ,i=1,\ldots ,N$. Each neuron $x_i$
is connected with other neurons $x_j$ by a random weighted coupling $J_{ij}$%
. Obviously, the connecting matrix ${\bf J}$ is asymmetric and the neural
systems exhibit complex spatial oscillations. As a simple version of coupled
neural systems designed by Zanette\cite{zanette 1998}, we use the parallel
dynamics for the updating neurons: 
\begin{equation}
\begin{array}{l}
x_i^1\left( t+1\right) =\left( 1-\varepsilon \right) \Theta \left(
h_i^1\left( t\right) \right) +\varepsilon \Theta \left( h_i^1\left( t\right)
+h_i^2\left( t\right) \right) \\ 
x_i^2\left( t+1\right) =\left( 1-\varepsilon \right) \Theta \left(
h_i^2\left( t\right) \right) +\varepsilon \Theta \left( h_i^1\left( t\right)
+h_i^2\left( t\right) \right)
\end{array}
\label{eq. 01}
\end{equation}
Here $h_i^k(t)$ is the local field of the $i$-th neuron and is expressed by 
\begin{equation}
h_i^k(t)=\sum\limits_{j=1}^NC_{ij}J_{ij}x_j\left( t\right)  \label{eq. 02}
\end{equation}
where $C_{ij}\in \left\{ 0,1\right\} $ is used to denote the linking status
between the $i$-th neuron and the $j$-th neuron. The dilution factor $C_{ij}$
is independent identically distributed random variable. It is selected by%
\cite{chen 2000} 
\[
{\rm if}\quad z\leq d,\ {\rm then}\quad C_{ij}=1;\quad {\rm else}\ C_{ij}=0 
\]
where $z\in [0,1]$ is a random number and $d\in [0,1]$ denotes the linking
degree of networks. The activation function $\Theta \left( r\right) $ is
defined as $\Theta \left( r\right) =\left[ 1+\tanh \left( \beta r\right)
\right] /2$. In there, $\beta \equiv 1/T$ characterizes a measure of the
inverse magnitude of the amount of noise affecting this neuron, acting as
the role of reciprocal of temperature in analogy to thermodynamic systems.
For convenience, we set $\beta =10$ through all simulations.

Obviously, the first term on the right-hand side of Eq. (\ref{eq. 01})
pictures the total response from its own internal units. The second term
expresses the interaction of the summation of the received signals from the
neurons with the same position in two networks. The parameter $\varepsilon
\in [0,1]$ named as coupling intensity, describes the interaction degree
between coupling subsystems. When the intensity $\varepsilon \ll 1$, it is
easy to see that the coupling sub-systems evolute independently. On the
other limit case, $\varepsilon \approx 1$, the coupling subsystems are
governed by the same dynamical law and will be synchronized very easily.

For measuring the coherence in the collective activity of the neural
systems, a time-dependent important feature $u_i\left( t\right)
=\sum\nolimits_{k=1}^Nx_k^i\left( t\right) $ for each network $i=1,2$ is
introduced. When the global coupling of two systems is absent, $\varepsilon
\approx 0$, the $u_i(t)$ will update independently and impossibly get to
synchronization since the initial conditions in subsystems are different. On
the other hand, the activity signals of two subsystems will be identical if
the coupling systems come to be synchronous. Figure $1\left( a\right) $
shows that the synchronization for this extend system successfully takes
place in $t=275$ for systems with $N=100$, while the coupling intensity $%
\varepsilon =0.34$ and the linking degree in each subsystem $d=0.2$.

To show the degree of synchronization in this coupling system, the
dispersion of activity patterns is defined as 
\begin{equation}
D\left( t\right) =\frac 12\sum\limits_{i=1}^2\sum\limits_{k=1}^N\left[
x_k^i\left( t\right) -\bar{x}_k\left( t\right) \right] ^2  \label{eq. 03}
\end{equation}
where $\bar{x}_k\left( t\right) =2^{-1}\sum\nolimits_{i=1}^2x_k^i\left(
t\right) $ denotes the average activity of neurons occupying the $k$-th
position in both subsystems at time $t$. Figure $1\left( b\right) $ shows
that the dispersion with a logarithmic scale evolves in time with the same
synchronous conditions in Fig. $1\left( a\right) $.

It is obvious that the larger coupling intensity $\varepsilon $ makes the
more easily synchronization arises for the system with the same other
parameters. Concomitantly, the case is whether there exists a critical
coupling parameter $\varepsilon _c$, and furthermore, whether there exists a
dependent relationship between $\varepsilon _c$ and the topological
structure in subsystems. In fact, since the evolution of networks is
sensitive to the varied initial status and the different random connecting
weight matrix in system with the same linking degree $d$, it is impossible
to find an identical value of $\varepsilon _c$. However, the fact that the
dispersion for these $\varepsilon _c$ corresponding to varied initial status
and connecting matrices from our simulations is small brings our notice to
investigate the qualitative curve of $\varepsilon _c$ {\it vs.} $d$.

Figure $2\left( a\right) $ shows a plot of the critical coupling intensity $%
\varepsilon _c$ versus the linking degree $d$ in subsystems with the size $%
N=200$. One can see that the qualitative relation between $\varepsilon _c$
and $d$ is close to a sigmoidal curve. The larger $d$ arise, the larger $%
\varepsilon _c$ become. This can be explained that the evolution of
subsystems with larger $d$ is more stable and it needs more powerful
coupling parameter to drive their evolutions into synchronization. In Fig. $%
2\left( b\right) $, we present the plot of simulation with the stepsize of
linking degree $\triangle d=0.001$ in the same conditions of Fig. $2\left(
a\right) $. Comparing both plots of simulations, the agreement is excellent
for the global tendency of the qualitative behavior of $\varepsilon _c$
versus $d$.

From Fig. $3$, it follows that it is more difficult to come into
synchronization with the increase of the size of subsystems. It is clear,
however, that there exists a homologous asymptotic behavior in the area of
larger linking degree. Now, the case is how the limit of coupling intensity
depends on varied size of networks. In Fig. $4$, we present the plot of $%
\varepsilon _c$ versus $1/N$ for the linking degree $d=1$ which is identical
to the limit case. The form of the limit coupling intensity as a function of
the inverse of size of networks calls for a fitting of these data with a
exponential decay function 
\begin{equation}
\varepsilon _c=A+Be^{-1/(C*N)}  \label{eq. 04}
\end{equation}
where the constant $A=0.44\pm 0.024$, $B=0.44\pm 0.020$ and $C=0.0066\pm
0.00071$. It follows that the maximal critical coupling intensity
corresponding to $N\rightarrow \infty $ is set as $0.88\pm 0.044$, and {\it %
vice versa}, the minimal $\varepsilon _c$ is $0.44\pm 0.024$ if both global
connecting subsystems designed by Eq. (\ref{eq. 01}) can be come to
synchronization.

Another important topic is the fraction of coupling neurons between two
subsystems. The considered coupling system can be viewed as a structure made
of two horizontal layers of networks. Apparently, from the definition Eq. (%
\ref{eq. 01}) of the above investigated systems, the neurons are involved in
global vertical coupling interactions between two layers, or the
dimensionality of coupling parameter is identical to the size of subsystems.
Considering the real physical systems or the potential applications, the
coupling interactions must be diluted and modified with time. As a result,
the systems defined by Eq. (\ref{eq. 01}) can be redefined as 
\begin{equation}
\begin{array}{l}
x_i^1\left( t+1\right) =\left( 1-\varepsilon \xi _i\left( t\right) \right)
\Theta \left( h_i^1\left( t\right) \right) +\varepsilon \xi _i\left(
t\right) \Theta \left( h_i^1\left( t\right) +h_i^2\left( t\right) \right) \\ 
x_i^2\left( t+1\right) =\left( 1-\varepsilon \xi _i\left( t\right) \right)
\Theta \left( h_i^2\left( t\right) \right) +\varepsilon \xi _i\left(
t\right) \Theta \left( h_i^1\left( t\right) +h_i^2\left( t\right) \right)
\end{array}
\label{eq. 05}
\end{equation}
where $\xi _i\left( t\right) \in \left\{ 0,1\right\} $ is a random number
with probability $1-p$ and $p$, respectively.

For revealing the association between the critical coupling fraction $p_c$
and the coupling intensity $\varepsilon $, the qualitative diagram of $p_c$
versus $\varepsilon $, is shown based on numerical simulations for systems
with $N=100$, $d=0.5$ (see figure $5\left( a\right) $). It is easy to get an
acceptable conclusion that $p_c$ decreases with the increase of $\varepsilon 
$ . Note that the series of turning points corresponding to $p_c=1$ in Fig. $%
5\left( b\right) $ are equivalent to the points in plot of $\varepsilon _c$
versus $d$ (cf. Fig. $3$).

In addition, another valuable informations about the minimal critical
fraction for synchronization of this extend system denoted by Eq. (\ref{eq.
05}) can be revealed from another critical point at $\varepsilon =1$ in Fig. 
$5\left( a\right) $ and Fig. $5\left( b\right) $. In Fig. $(6)$, the minimal
critical coupling probability for various linking degree of subsystem with $%
N=400$ is presented. During the evolution of networks, the neuron of each
site in both networks updates due to the competing effect of the local rules
and the coupling mechanism. It is clear that the region above the curve is
the synchronization part, while the lower part is desynchronizatious, thus
the curve embodies a competing relationship between local correlation and
stochastic coupling. It is possible to give the minimal critical coupling
fraction for various linking degree in synchronization of this extended
system with $N\rightarrow \infty $ , which is more analogous to the case of
real biotic systems, through analyzing curves of $p_c$ {\it vs.} $d$ for
various size of subsystem. However, considering our computational device,
the more intensive and detail work is left out in there.

In this paper, we have studied the critical features of coupling parameter
in the synchronization of neural networks for various structural topology.
We obtain the exponential decay form in the case of global coupling among
subsystems and fully linking in each network. We find that it exists the
maximal and minimal critical coupling intensity for synchronization in this
extend systems. For the case of partial coupling, a primary result about the
critical coupling fraction for various linking degrees of networks is shown.
Considering the definition of our model is analogous to coupled map lattice,
it is easy to generalize the present work to other extended systems, such as
coupled ordinary differential equations and partial differential equations.

\medskip

This work was supported by the Doctoral Research Foundation awarded by
Lanzhou University and Innovation Project of CAS with Grant No. kzcx1-09. We
wish to thank Prof. H. Zhao for the most constructive and fruitful
discussions.

\bigskip

FIG. 1. The synchronization of diluted networks with $N=100$, $\varepsilon
=0.34$ and $d=0.2$. (a) The evolution of time-dependent activity of both
subsystems comes into synchronization at $t=275$. (b) logarithm of
dispersion of both networks.

\medskip

FIG. 2. The simulations of relationship between the critical coupling
parameter $\varepsilon _c$ and the linking degree $d$ with $N=200$,
corresponding to stepsize of linking degree (a) $\triangle d=0.05$ and (b) $%
\triangle d=0.001$.

\medskip

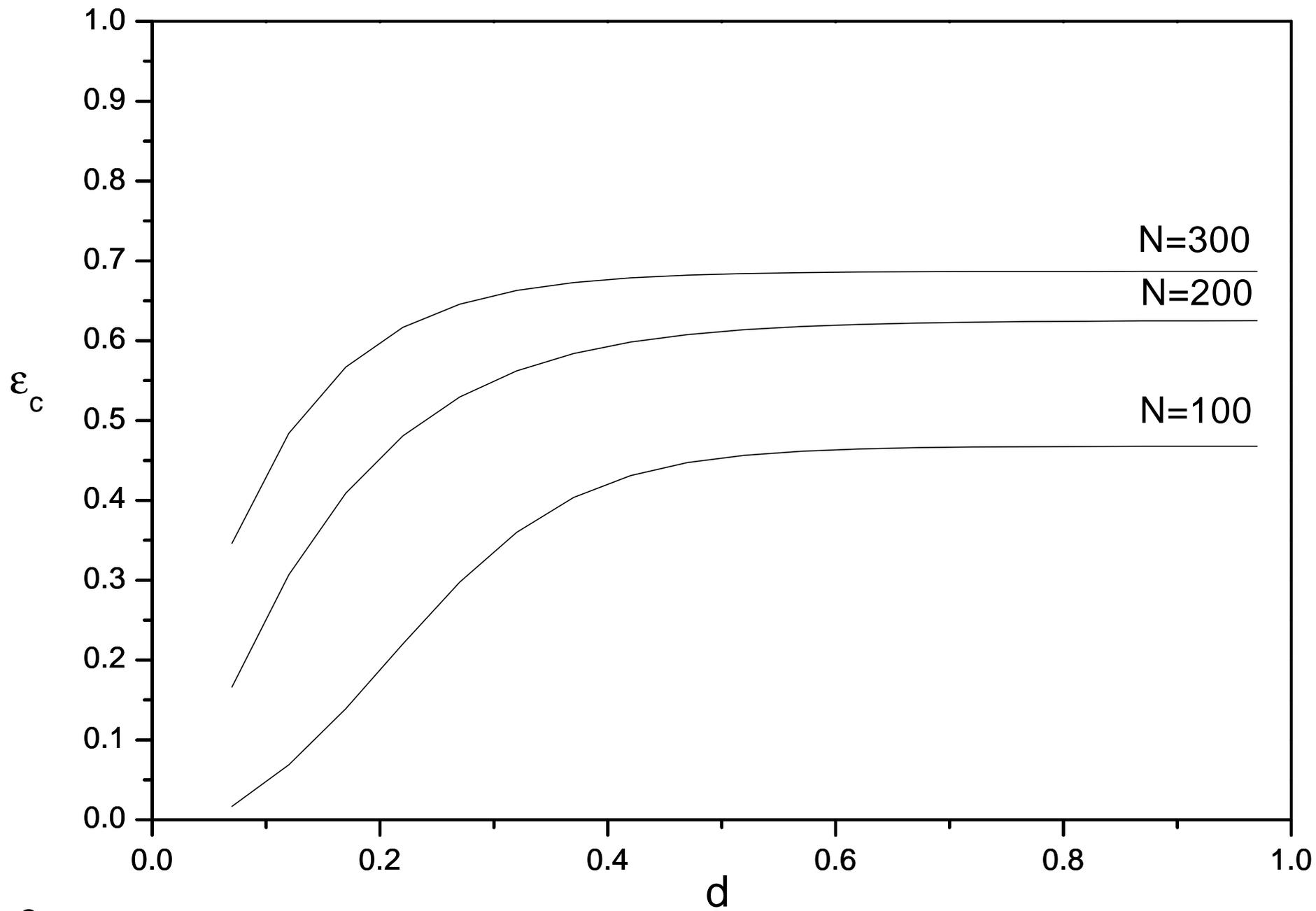
FIG. 3. The qualitative relationship of $\varepsilon _c$ versus $d$ for
varied size of networks $N$.

\medskip

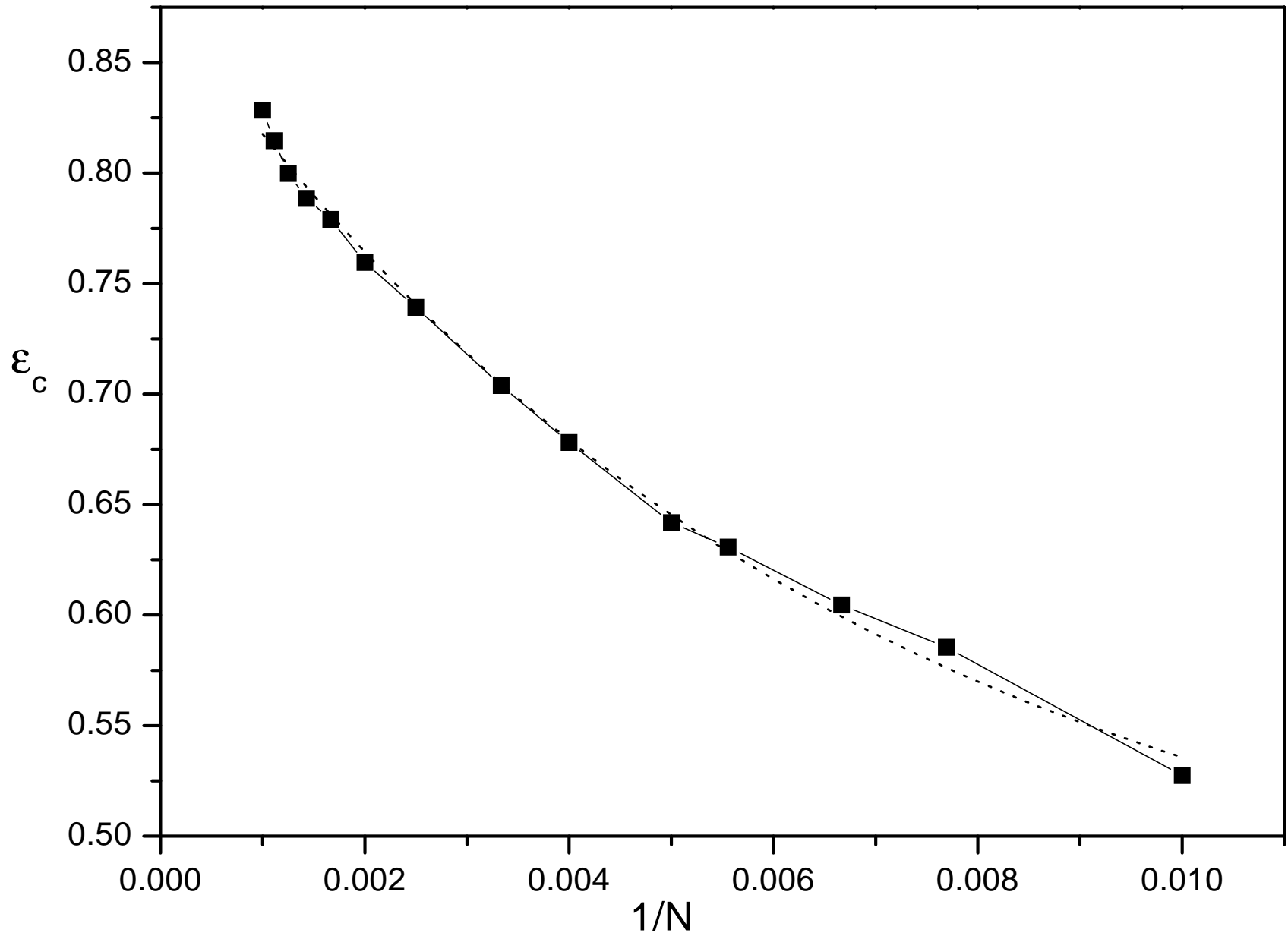
FIG. 4. The plot of $\varepsilon _c$ versus $1/N$ for varied size of
networks in the limit case $d=1$. The exponential decay fitted curve of this
relationship is shown by a dotted line.

\medskip

FIG. 5. The synchronization diagram of minimal coupling probability $p_c$
versus the corresponding coupling intensity $\varepsilon $, (a)for systems
with size $N=100$, $d=0.5$ (b) for systems with varied linking degree $d$
and $N=100$ in the forms of qualitative curves.

\bigskip 

FIG.6. The qualitative relationship of $p_c$ versus $d$ with the subsystem
size $N=400$.

\medskip

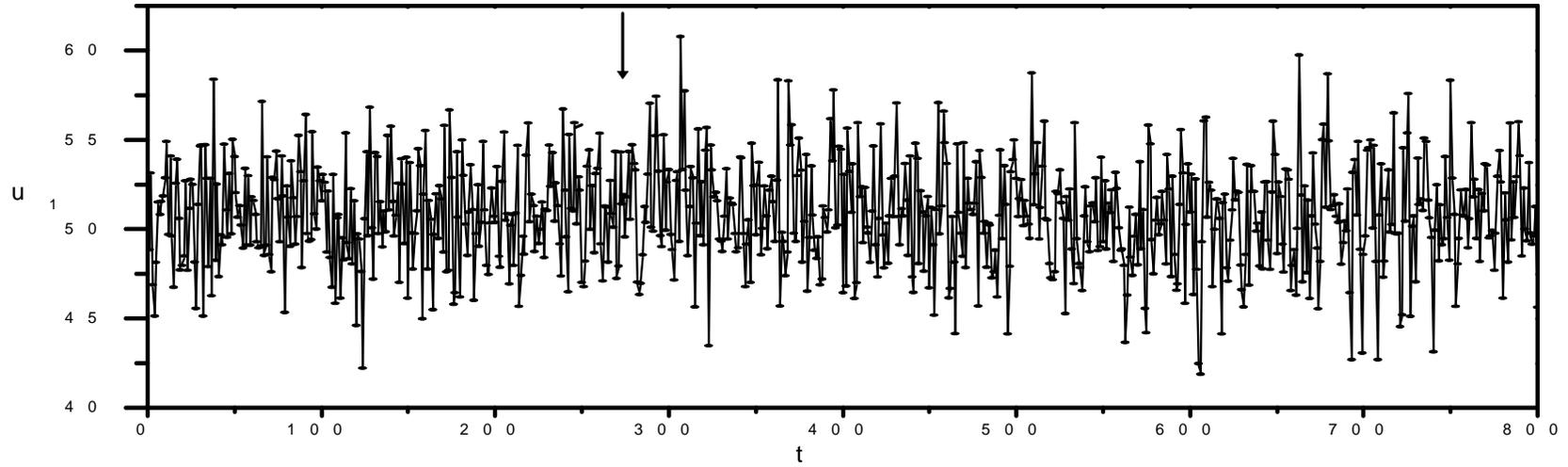
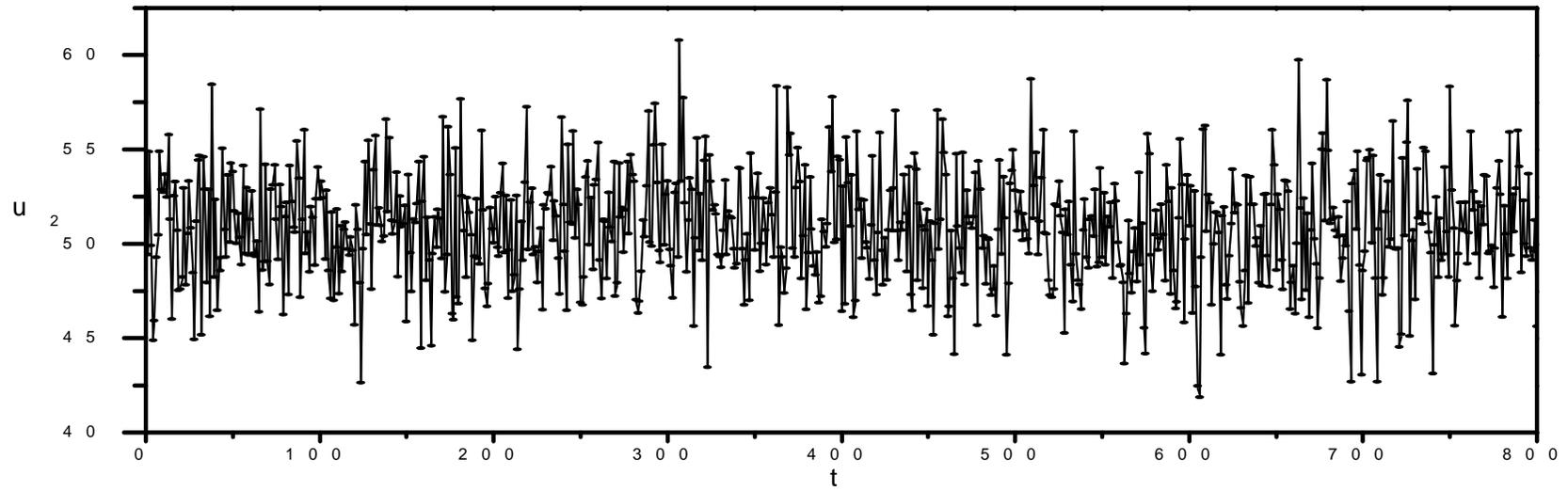
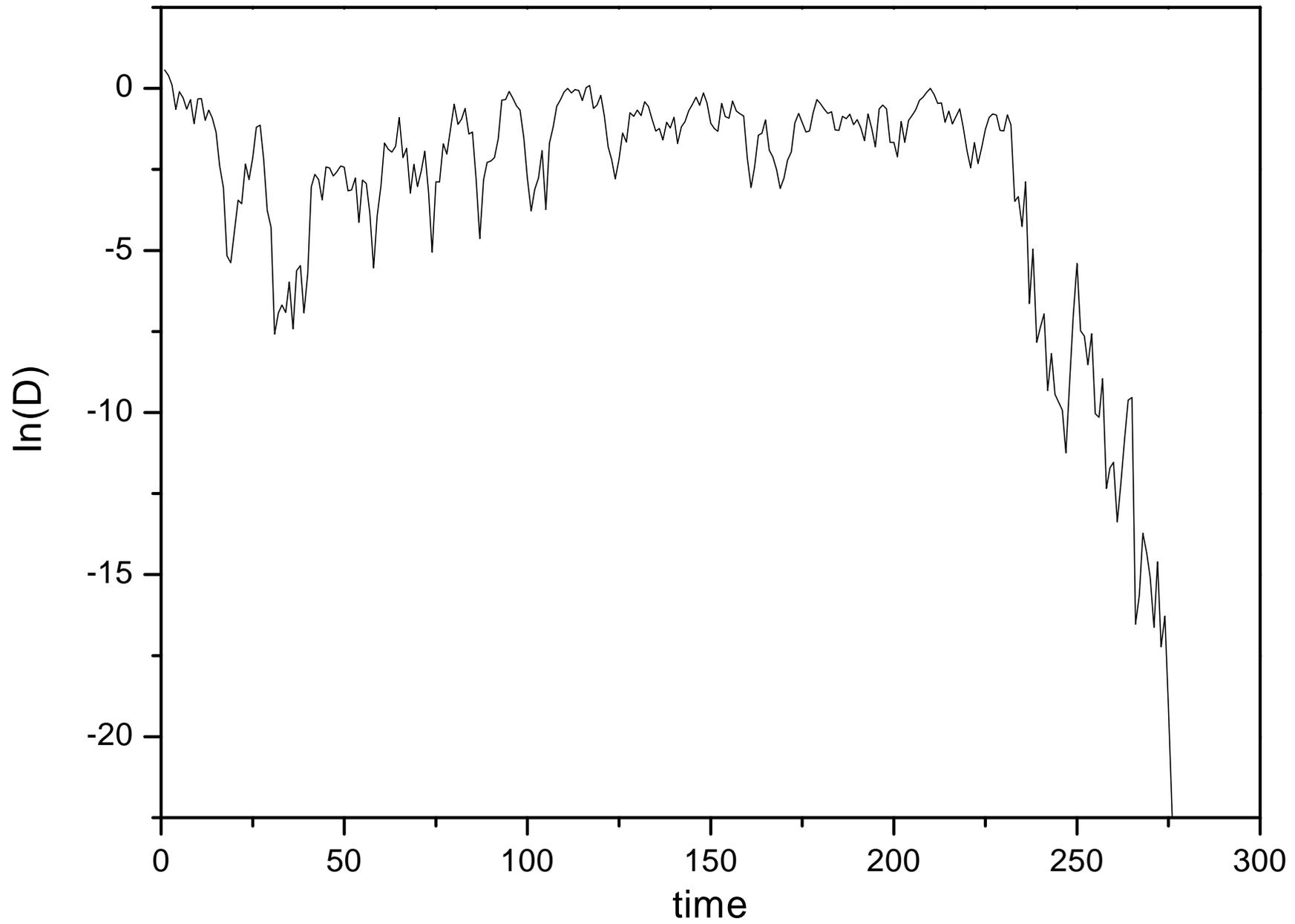
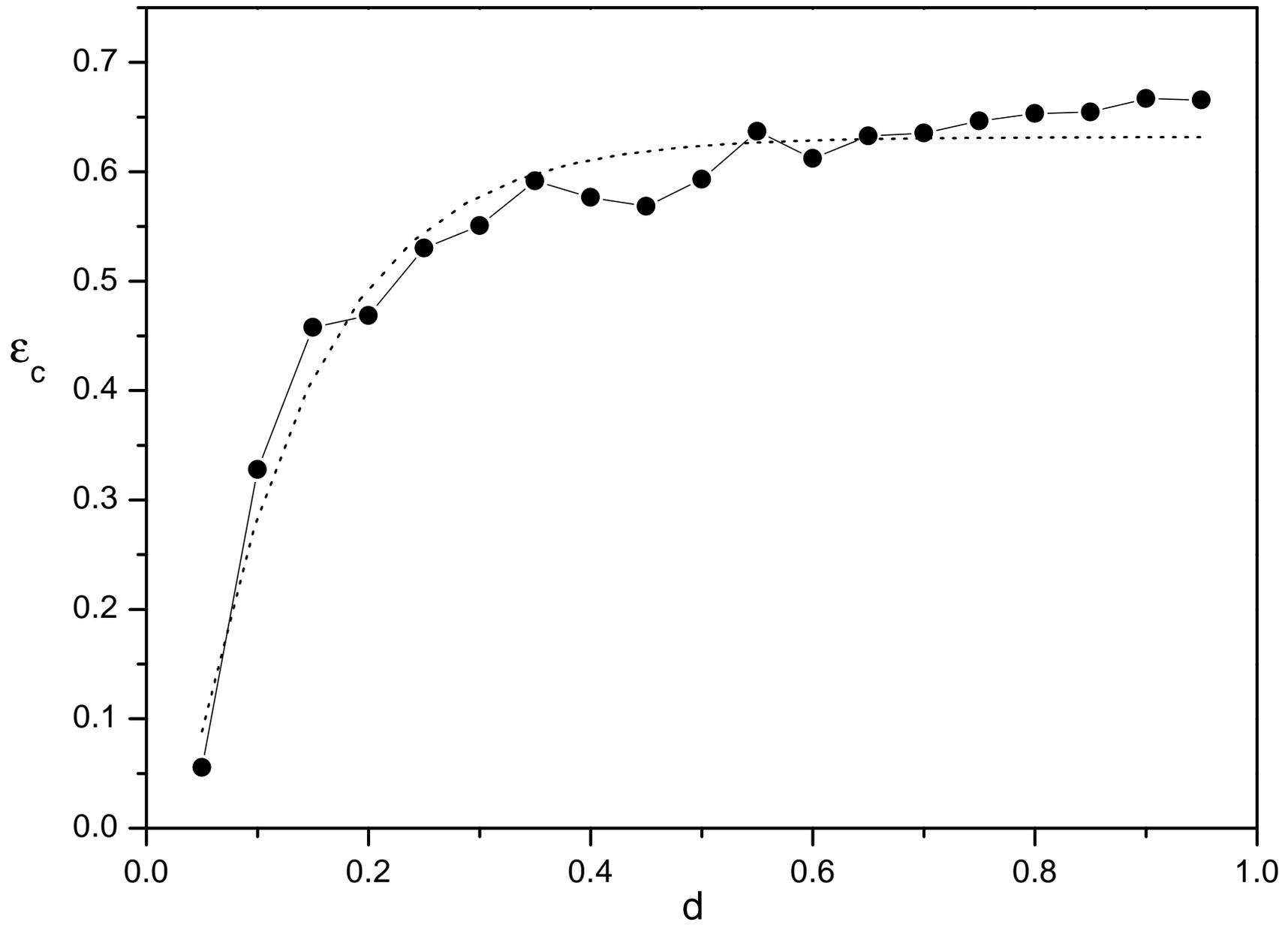
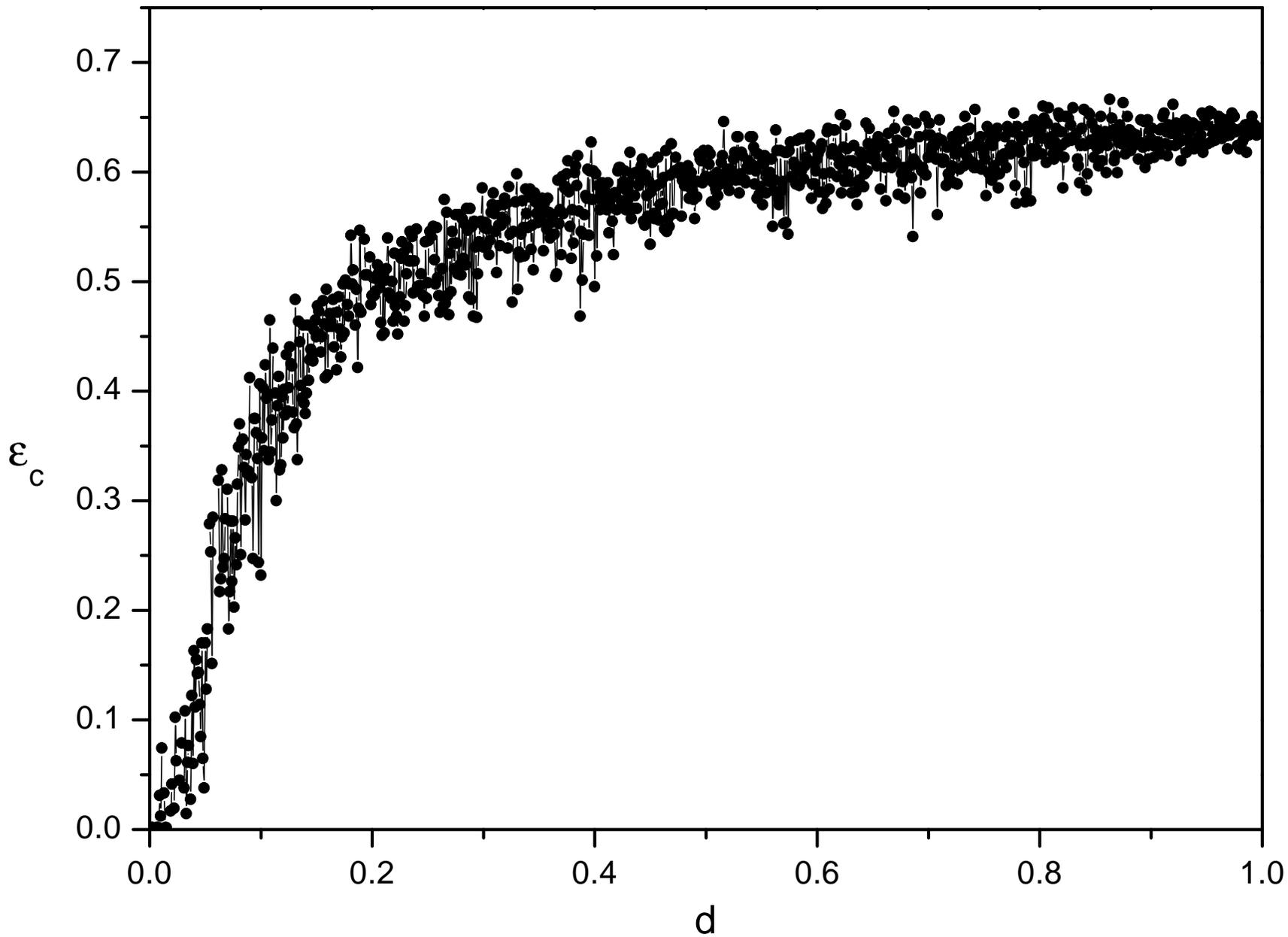
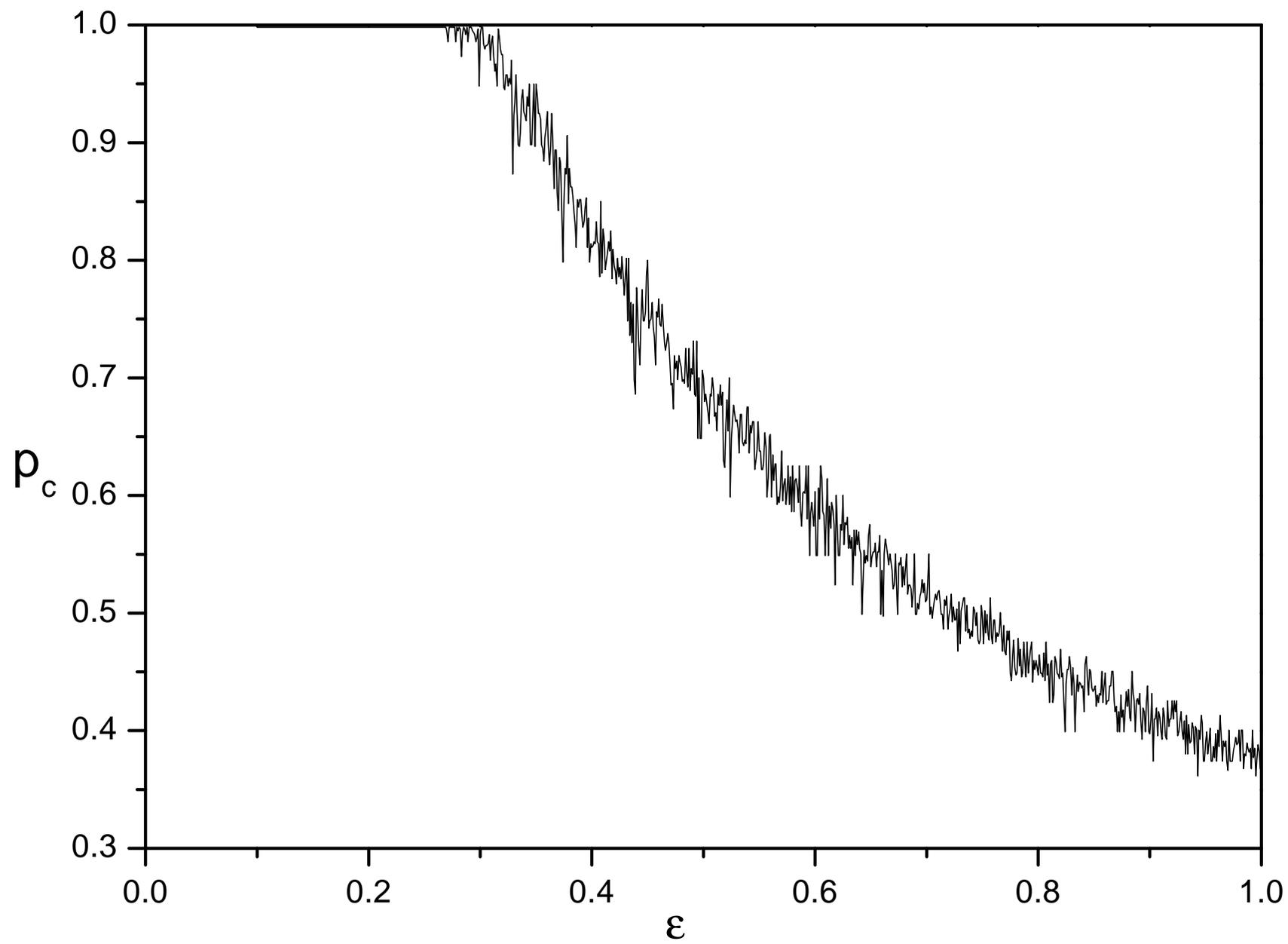
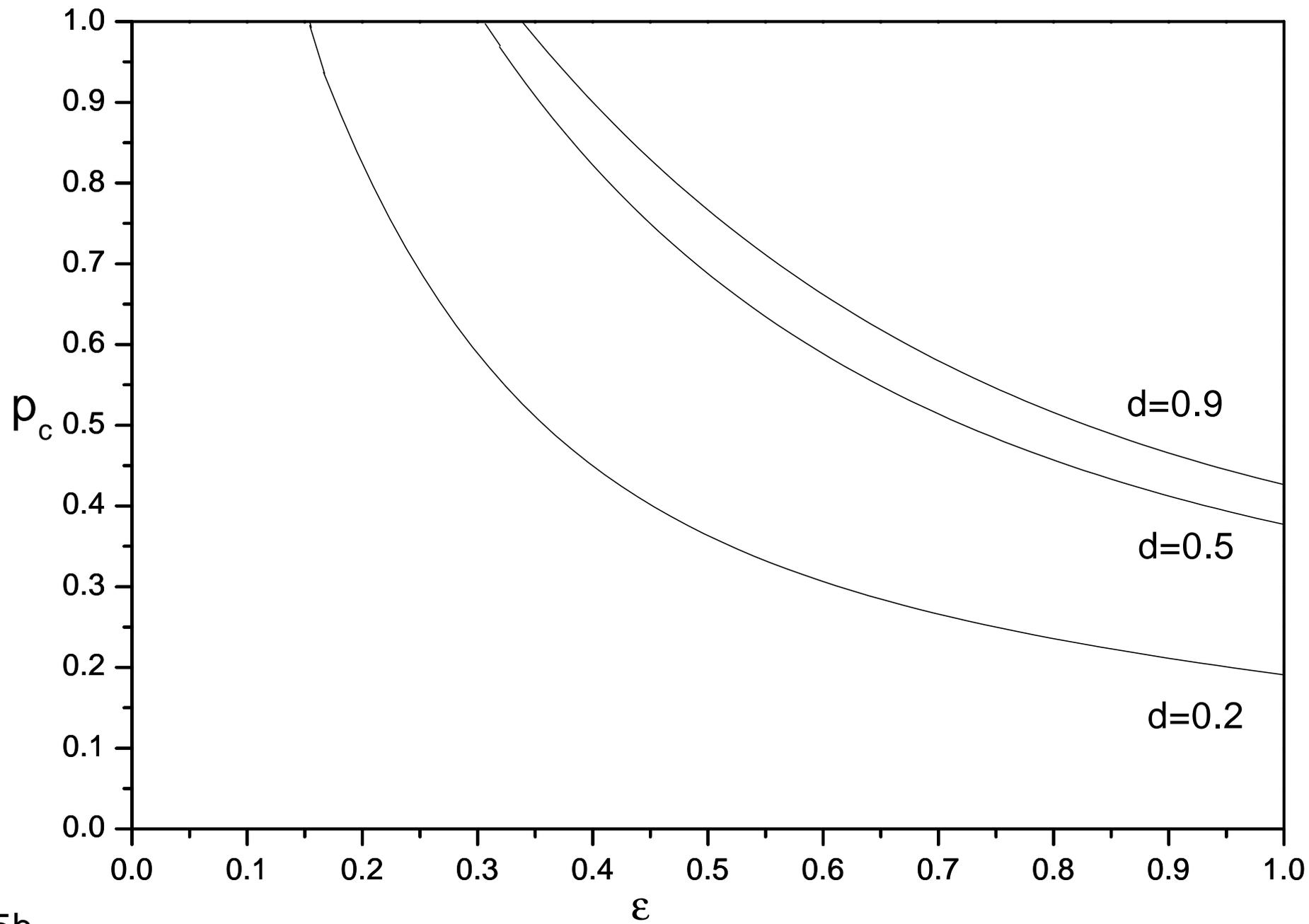
\end{document}